\documentclass[12pt, preprint]{aastex}
\usepackage{natbib}
\usepackage{amsmath}
\usepackage{graphicx}

\shorttitle{Gamma Rays from Star Formation in Clusters}

\shortauthors{Storm, Jeltema \& Profumo}

\begin{document}

\title{Gamma Rays from Star Formation in Clusters of Galaxies}

\author{Emma M. Storm\altaffilmark{1}, Tesla E. Jeltema\altaffilmark{1,2} and Stefano Profumo\altaffilmark{1,2}}

\altaffiltext{1}{Department of Physics,  University of California, 1156 High St., Santa Cruz, CA 95064, USA}
\altaffiltext{2}{Santa Cruz Institute for Particle Physics,  University of California, 1156 High St., Santa Cruz, CA 95064, USA}

\begin{abstract} 
Star formation in galaxies is observed to be associated with gamma-ray emission, presumably from non-thermal processes connected to the acceleration of cosmic-ray nuclei and electrons. The detection of gamma rays from starburst galaxies by the \textit{Fermi} Large Area Telescope (LAT) has allowed the determination of a functional relationship between star formation rate and gamma-ray luminosity \citep{Bechtol2011}. Since star formation is known to scale with total infrared ($8-1000$ $\mu$m) and radio ($1.4$ GHz) luminosity, the observed infrared and radio emission from a star-forming galaxy can be used to quantitatively infer the galaxy's gamma-ray luminosity. Similarly, star forming galaxies within galaxy clusters allow us to derive lower limits on the gamma-ray emission from clusters, which have not yet been conclusively detected in gamma rays. In this study we apply the functional relationships between gamma-ray luminosity and radio and IR luminosities derived in \citet{Bechtol2011} to a sample of the best candidate galaxy clusters for detection in gamma rays from \citet{Ackermann2010c} in order to place lower limits on the gamma-ray emission associated with star formation alone in galaxy clusters.  We find that several clusters have predicted gamma-ray emission from star formation that are within an order of magnitude of the upper limits derived in \citet{Ackermann2010c} based on non-detection by \textit{Fermi}-LAT. Given the current gamma-ray limits, star formation likely plays a significant role in the gamma-ray emission in some clusters, especially those with cool cores. We predict that both \textit{Fermi}-LAT over the course of its lifetime and the future Cherenkov Telescope Array will be able to detect gamma-ray emission from star-forming galaxies in clusters.
\end{abstract}


\section{INTRODUCTION}

Galaxy clusters are the most massive gravitationally bound objects in the universe, and contain a dense population of galaxies surrounded by gas distributed throughout the intracluster medium (ICM). The complex environment of galaxy clusters is bound to host, at some luminosity level, processes that lead to the production of gamma rays. In particular, a significant fraction of the gamma-ray emission is thought to be associated with the ICM, and more specifically with cosmic ray (CR) populations accelerated by shocks and turbulence within the ICM, as well as, possibly, with dark matter annihilation and decay (first suggested in \citealt{Totani2004}). In the ICM, CR protons and other nuclei collide with dust and gas particles and decay to charged and neutral pions, which in turn decay into leptons and gamma rays, respectively. CR electrons, if sufficiently energetic, can inverse-Compton scatter photons from the Cosmic Microwave Background up to gamma-ray energies (see e.g., \citealt{Rephaeli2008} for a general description of nonthermal activity in the ICM of clusters of galaxies).

Clusters have not, however, been detected yet as gamma-ray sources. The Large Area Telescope (LAT), the primary instrument aboard the \textit{Fermi} Gamma-ray Space Telescope, has placed upper limits on the flux from the best candidate clusters and used these upper limits to place constraints on the cosmic ray populations in clusters \citep{Ackermann2010c}. Atmospheric Cherenkov telescopes such as H.E.S.S. and MAGIC have also reported null results from observations of selected clusters of galaxies, such as Perseus \citep{Aleksic2012}, Coma \citep{Aharonian2009}, and A0085 and A0496 \citep{Aharonian2009a}.

In addition to the gamma rays produced by the ICM, the cluster member-galaxies themselves are at some level a source of gamma rays. Ordinary galaxies such as our own Milky Way, its satellite galaxies the Large Magellanic Cloud (LMC) and Small Magellanic Cloud (SMC), and the nearby Andromeda Galaxy (M31) have all been detected in gamma rays (MW: \citealt{Abdo2009}, LMC: \citealt{Abdo2010h}, SMC: \citealt{Abdo2010g}, M31: \citealt{Abdo2010i}). In addition, the recent detection by the \textit{Fermi}-LAT of four star-forming galaxies, M82, NGC 253 \citep{Abdo2010c} and NGC 1068 and NGC 4945 \citep{Lenain2010}, allowed for the determination of a quantitative functional relationship between the star formation rate and the gamma-ray luminosity \citep{Bechtol2011}. 

While gamma-ray emission has not yet been detected from galaxy clusters, some minimum gamma-ray emission resulting from star formation activity in cluster members must exist. Lower limits on the gamma-ray flux from clusters of galaxies can therefore be determined by considering only the emission from cluster members with ongoing star formation, and can also provide insight into the star-forming population of galaxies within clusters, as compared to the field. This is the scope of the present study. The relations derived in \citet{Bechtol2011} are used to predict lower limits on the gamma-ray luminosity coming from star formation within cluster members alone, provided total IR and/or radio luminosity measurements for a sample of clusters.  We then compare these estimates to the \textit{Fermi} upper limits on the gamma-ray luminosity from the same clusters \citep{Ackermann2010c}, and assess the potential of the \textit{Fermi}-LAT over its lifetime as well as of the future Cherenkov Telescope Array for detection of gamma-rays from star-forming galaxies in clusters. 

In the following section, we review the sources of multiwavelength emission resulting from star formation in galaxies and clusters.  In section 3, we describe our cluster sample selection and the available IR and radio data. In section 4 we describe our results and present lower limits on the gamma-ray emission from a selection of clusters. We use the lower limits to explore the possibility of detection by various gamma-ray telescopes in section 5, and conclude in section 6.

\section{EMISSION FROM STAR FORMATION}

The bulk of the bolometric luminosity of young stars is emitted in the ultraviolet (UV), which is then efficiently reprocessed into infrared (IR) light by dust in the interstellar medium (ISM) within galaxies such that their spectral energy distributions are peaked in the IR \citep{Sanders1996}. The total IR luminosity ($8-1000 \mu$m) thus traces closely the rate of star formation in a galaxy and is an extinction-free way to measure the star formation rate (SFR) since the ISM is optically thin to IR \citep{Kennicutt1998}. However, the total IR luminosity is not directly measured; rather, single-band measurements (24 $\mu$m and 60 $\mu$m are commonly employed wavelengths) are converted to total luminosities either integrating over template SEDs (e.g., \citealt{Dale2002}) or using relations derived from modeling the dust emission in the ISM (e.g., \citealt{Sanders1996}). Both techniques are common in the literature and a determination of the total IR luminosity thus depends somewhat on the particular templates or relations used for conversion. Discrepancies between methods developed specifically by \citet{Dale2002} and \citet{Sanders1996} vary from $\sim4\%$ for starburst galaxies, which have high SFRs and generally are very IR-bright (L$_{8-1000 \mu \textrm{m}}>10^{11}$L$_{\sun}$), to $\sim25\%$ for low-luminosity galaxies, with \citet{Sanders1996} method predicting smaller values for L$_{8-1000 \mu \textrm{m}}$ than those predicted by \citet{Dale2002}.

Massive stars typically end their life-cycles as supernovae, whose remnants (SNRs) are likely accelerators of cosmic ray protons and electrons (e.g., \citealt{Volk1989} and references therein). The CR electrons interact with magnetic fields within a galaxy and quickly lose their energy through synchrotron radiation in the radio continuum. Measured at $1.4$ GHz, the radio luminosity is tightly correlated with total IR luminosity in star-forming galaxies over several orders of magnitude and is therefore also used as an indicator for SFR  \citep{Condon1992, Yun2001}. This correlation can be explained by a simple model that treats galaxies as ``calorimeters'' of both UV photons and CR electrons \citep{Volk1989}. CR electrons lose most of their energy through synchrotron radiation and UV photons lose most of their energy through absorption and re-emission by dust in the ISM. In this model, if the ratio of synchrotron emission versus total energy losses in the CR electron population is large, then the radio luminosity becomes a proxy for SFR. In addition, if UV photons emitted by stars are effectively reprocessed into the IR, then IR luminosity would also trace with SFR. The L$_{1.4 \textrm{GHz}}$--L$_{8-1000 \mu \textrm{m}}$ ratio, which should be constant if both are due mainly to star formation, can be thus be used as an independent indicator of the SFR in galaxies. Active galactic nuclei (AGN), which can have large luminosities in both IR and radio due to processes other than just star formation, and which presumably also contribute to the gamma-ray emission in some clusters, deviate from this IR-radio correlation \citep{Yun2001}. We discuss this caveat in detail for the cluster sample under consideration here. 

Gamma rays are expected to scale with SFR in normal, star-forming galaxies, and are thought to arise mostly from CR nucleon collisions with the ISM and subsequent decay into charged and neutral pions, which then decay to leptons and gamma rays, respectively. At energy thresholds higher than those required for pion decay the gamma-ray spectrum is the same shape as the underlying CR nuclei population, which is controlled by the number of SNRs. Therefore the amplitude of the gamma-ray spectrum can be used as a proxy for SFR. This is supported by observations of the LMC in gamma rays, which has been spatially resolved by \textit{Fermi}-LAT. The LMC's star forming region 30 Doradus is associated with strong diffuse gamma-ray emission, which implies the CR proton intensity is also strongest in that region \citep{Abdo2010h}. The secondary CR electron population, resulting from charged pion decay, also contributes to the gamma-ray spectrum primarily by inverse-Compton up-scattering of starlight \citep{Bechtol2011}. 

Early in the mission, in all-sky survey mode, \textit{Fermi}-LAT detected two starburst galaxies, M82 and NGC 253 \citep{Abdo2010c}. Those starbursts were also detected in the very-high-energy regime by ground-based gamma-ray telescopes (M82 by VERITAS; \citealt{Collaboration2009} and NGC 253 by H.E.S.S.; \citealt{Acero2009}). As first reported in \citet{Lenain2010} and later in \citet{Bechtol2011}, \textit{Fermi}-LAT detected two additional starbursts, NGC 1068 and NGC 4945, which host AGN. \citet{Bechtol2011} examined a sample of 69 galaxies, mostly consisting of starbursts for which \textit{Fermi}-LAT has upper limits, 9 of which are associated with AGN. The sample includes the four starbursts detected by \textit{Fermi}-LAT and several detected local group galaxies which are not starbursts: the Small and Large Magellanic Clouds, the Milky Way, M31 and M33. \citet{Bechtol2011} found that gamma-ray luminosity scales with total IR and radio luminosities and hence with SFR across several orders of magnitude in luminosity and SFR. The relationships found between SFR and gamma-ray luminosity are power-law in nature and differ slightly depending on whether AGN are included in the sample, but the dispersion in the relationships dominate over the differences in the fits. 

Star formation in cluster galaxies occurs at slower rates compared to that in field galaxies (see e.g., \citealt{Reddy2004}, and references therein). However, the processes that suppress star formation in clusters are not well-understood. \citet{Chung2011} used infrared data from WISE (Wide-field Infrared Survey Explorer) to estimate SFRs for a sample of local clusters with known masses and found no correlation between cluster mass and SFR. This implies that clusters processes that scale with mass perhaps only weakly affect star formation, if at all. Ram pressure in particular scales up with cluster mass and is thought to influence star formation.  Ram pressure strips gas from a galaxy as it moves through the ICM, removing the raw material needed for star formation. Rates of star formation in cluster galaxies get smaller as the radial distance from the cluster center decreases (\citealt{Chung2011} and references therein) which supports the theory that ram pressure stripping strongly influences star formation. In addition, ram pressure is shown to decrease star formation in clusters in simulations \citep{Book2010} and in observations of the Virgo cluster \citep{Vollmer2008}. However, this conflicts with the lack of evidence for a correlation between mass and star formation in local clusters, described by \citet{Chung2011} and others (e.g., \citealt{Goto2005}). Galaxy-galaxy mergers and interactions also tend to trigger bursts of star formation in clusters, but are most common in poor clusters and groups, (e.g., \citealt{Martig2008}), while the best clusters for detection in gamma rays tend to be rich and massive. Still many cluster galaxies do show evidence of ongoing star formation, and the density of galaxies within clusters mean that the summed contribution to the gamma-ray emission may be significant.

Cool-core clusters display elevated SFRs in their central galaxies, as compared to non-cool-core clusters \citep{McDonald2011}. These clusters have high density, low temperature cores which suggest that cool gas is steadily flowing into the center (for a review, see \citealt{Fabian1994}). Therefore the elevated SFR may be due to this mass deposition onto the central galaxy, which would enhance star formation by supplying the brightest cluster galaxy (BCG) with a steady stream of raw material. \citet{Egami2006}, however, found only a weakly suggestive relationship between mass deposition rate and SFR in a sample of X-ray selected clusters.  In addition to several clusters of galaxies, in this work we also consider the possible gamma-ray emission from star formation in 3 BCGs located in cool-core clusters.

\section{METHODS}
We estimate here the gamma-ray luminosity associated with star formation in cluster galaxies from the total infrared luminosity ($8-1000$ $\mu$m) and from the $1.4$ GHz radio continuum luminosity using the relationships in \citet{Bechtol2011}. We restrict our analysis to emission from individual galaxies where there is active star formation, since star formation in clusters only occurs within galaxies. We performed a literature search for IR and radio data for cluster galaxies starting with the list of clusters in \citet{Ackermann2010c}, which places upper limits on the gamma-ray flux for the best candidate clusters based on nondetection by \textit{Fermi}-LAT. The clusters in \citet{Ackermann2010c} were selected from the HIFLUGCS catalog of brightest X-ray clusters \citep{Reiprich2002}. The clusters with the highest mass-to-distance-squared ratios were selected for analysis. Additionally, several clusters with nonthermal radio emission were included in the list of \citet{Ackermann2010c}. We performed a literature search for radio and IR data on all of the clusters in that list, excluding several that were close to the galactic plane or that had high redshifts (with the exception of the Bullet Cluster). We excluded AGN from our calculated luminosities, since AGN can be bright in IR and radio but the emission is not due to active star formation. The results of the literature search are presented in Table 1 and the corresponding estimated gamma-ray luminosities are summarized in Table 2; the upper limits provided by \citet{Ackermann2010c} are also included in Table 2. 

In looking for IR and radio data, we preferentially searched first for total IR ($8-1000$ $\mu$m) and $1.4$ GHz luminosity functions (LFs) of clusters with Schechter fits. We found fitted total IR LFs for two clusters, Coma \citep{Bai2006} and the Bullet Cluster \citep{Chung2010}. While we found a radio LF for Coma \citep{Miller2009} and several far-IR ($100-500$ $\mu$m) LFs for Virgo \citep{Davies2010}, they were not fitted and therefore we were unable to convert to total luminosities.

We expanded the search to include lists of luminosities of individual cluster members. \citet{Reddy2004} presented L$_{1.4 \textrm{GHz}}$ and L$_{60 \mu\textrm{m}}$ for members in 7 clusters. \citet{Rieke2009} found a tight correlation between L$_{60 \mu\textrm{m}}$ and L$_{8-1000 \mu\textrm{m}}$ and derived a relationship for converting one to the other; we used this relation to convert the 60 $\mu$m data in \citet{Reddy2004} to total IR. We converted each radio and IR galaxy luminosity to gamma-ray using the relationships in \citet{Bechtol2011} that exclude AGN, then summed up the gamma-ray luminosities of individual galaxies to obtain a total cluster gamma-ray luminosity. We excluded galaxies classified as Seyferts or LINERs, which are typically AGN, from our calculations. Any galaxies not identified as Seyferts or LINERs that had one anomalously large luminosity (radio or IR) as compared to the other were checked in the literature, and those identified as hosting an AGN were also excluded. In \citet{Reddy2004}, only cluster members detected in both radio and IR were included, so total cluster luminosities using that data are most likely underestimates in both IR and radio. In the cases for which we had luminosity functions, we first integrated the LF, then converted this total cluster luminosity to gamma-ray. Since the relationship between IR and gamma-ray luminosity is slightly steeper than linear, this procedure results in an overestimate of the true total gamma-ray luminosity by $\sim20\%$. This, however, is smaller than the intrinsic dispersion in the relationship reported in \citet{Bechtol2011}, which we take as the uncertainty in our estimates of  gamma-ray luminosities.

Uncertainties in galaxy IR and radio luminosities were not reported in the papers we found. We therefore report the uncertainties in the estimated gamma-ray luminosities as the dispersion in the power-law relationships derived in \citet{Bechtol2011}. For L$_{0.1-100 \textrm{GeV}}$ estimated from L$_{8-1000 \mu\textrm{m}}$, the uncertainty is 0.25 dex (excluding AGN). For L$_{0.1-100 \textrm{GeV}}$ estimated from L$_{1.4 \textrm{GHz}}$, the uncertainty is 0.19 dex (excluding AGN). While \citet{Bechtol2011} provides uncertainties in the fit parameters, the scatter in the L$_{0.1-100 \textrm{GeV}}$-L$_{8-1000 \mu\textrm{m}}$ and L$_{0.1-100 \textrm{GeV}}$-L$_{1.4 \textrm{GHz}}$ relationships dominate over the fit parameter uncertainties.

In cool core clusters, there is evidence for active star formation in the central cluster galaxies \citep{McDonald2011}. We found radio fluxes for the brightest cluster galaxy that are not associated with AGN in 3 cool core clusters, Ophiuchus, A2029, and A2142, and included them in our sample. While Ophiuchus is close to the galactic plane, we include it as a representative cool-core cluster. The gamma-ray luminosities estimated from these clusters are most likely severe underestimates, but it is interesting to note that for cool core clusters the BCG alone can have a significant predicted gamma-ray emission due to increased star formation.

We note that there is observed large scale ($\sim1$ Mpc) diffuse radio emission from the ICM of some galaxy clusters which indicate the existence of cosmic rays that may also be responsible for as-yet-undetected gamma-ray emission. \citep{Ferrari2008}. These ICM cosmic rays are accelerated primarily by shocks and turbulence in the ICM, as opposed to supernova remnants in cluster galaxies, and are not related to star formation in galaxies. Observations of clusters in the IR are always associated with individual galaxies (see e.g., \citealt{Coppin2011}). We thus choose to consider only the radio and IR emission from galaxies themselves in order to place conservative lower limits on the gamma-ray emission from clusters from star formation in cluster galaxies alone.

\section{RESULTS}

We present our results in Table 1 and 2 and in the associated Figures 1 and 2. Table 1 shows the IR and radio luminosities for the clusters under consideration, while Table 2 presents our calculation for the gamma-ray luminosity lower limits, and compares the latter with the upper limits from \citet{Ackermann2010c}. Finally, Figures 1 and 2 show the calculated gamma-ray luminosities as a function of the IR and radio luminosities, respectively. 

For several clusters the lower limits on gamma-ray luminosity predicted from star formation are within an order of magnitude or so of the upper limits derived in \citet{Ackermann2010c}. The lower limit luminosity of the BCG (IC 1101) in the cool-core cluster A2029 in particular is quite close to the upper limit. The core of A2029 shows extended radio emission which suggests it hosts an AGN \citep{Taylor1994}. A2029 has also been observed in X-rays, but it is unclear whether the X-ray emission is due to an AGN since the emission is not point-source-like and does not fit a power-law spectrum, as is typical for X-ray AGNs \citep{Clarke2004}. Additionally, the star formation rate calculated nominally from the presented radio luminosity in \citet{McDonald2011} using the relationship in \citet{Kennicutt1998} yields an SFR that is higher than that of any local starburst galaxy; therefore is it most probable that only a fraction of the radio luminosity is due to star formation. However, the large ratio of far UV to H$\alpha$ emission suggests that the BCG of A2029 is an older starburst galaxy with ongoing star formation activity \citep{McDonald2011}, so it is included in our sample.

The Bullet Cluster may not be detectable by \textit{Fermi}-LAT due to its distance ($z\simeq0.3$). It is however an interesting cluster, with a total IR LF available in the literature and a high IR luminosity, and was therefore included in our sample. If a cluster similar in mass and activity to the Bullet existed nearby it would likely be observed in gamma rays. The Bullet Cluster's predicted emission from star formation is much lower than the upper limit on its gamma-ray luminosity given its distance, but this lower limit on the Bullet's gamma-ray luminosity is comparable to the detection threshold for a closer cluster with similar mass such as Coma (although the IR LF is a different shape than that of Coma \citep{Chung2010}).

We present two total IR luminosities for the Coma cluster, one calculated from the total IR luminosity function \citep{Bai2006} and the other from a sum of individual member luminosities \citep{Reddy2004}. The galaxy sample used to calculate the LF in \citet{Bai2006} does include 3 AGN, but their contribution to the total luminosity is negligible. We therefore used the relation in \citet{Bechtol2011} that excludes AGN, as for all other clusters. As expected, the gamma-ray luminosity as calculated from the IR luminosity function is larger than the gamma-ray luminosity calculated from summing individual galaxy luminosities.

\section{DISCUSSION}

\subsection{Detections with \textit{Fermi}-LAT}

Given the lower limits presented in the previous section, we examine here which clusters, if any, may be detectable by \textit{Fermi}-LAT over the course of the instrument's lifetime. \textit{Fermi}-LAT is an all-sky survey telescope, covering the full sky every 2 orbits, or 3 hours. It was launched in June 2008 and is currently funded through at least 2016; it is expected to be operational for a total of 10 years. Its sensitivity gets better over time, and if we assume a 10-year flux sensitivity of $\sim 3\times 10^{-9}$ cm$^{-2}$ s$^{-1}$ for detecting a point source with photon index $\alpha_{\gamma}=2.2$ at the 5$\sigma$ level, which is the best fit value for star-forming galaxies from \citet{Bechtol2011}, then several clusters have lower limit fluxes that are above or within a factor of 2-3 of this sensitivity limit, including A2029, Virgo, Coma, A1367, and Hydra, with the flux of A2029 strictly over the $3\times 10^{-9}$ cm$^{-2}$ s$^{-1}$ limit. The photon index $\alpha_{\gamma}=2.2$ is chosen as it is the value used in \citet{Bechtol2011} to predict upper limits on gamma-ray luminosities from non-detected starburst galaxies. However, Virgo is a large, extended object, and it hosts a bright AGN, M87, which may make it difficult to detect emission in excess of this gamma-ray point source. Recently, \citet{Han2012} claimed a detection of diffuse gamma-ray emission from the central $3^{\circ}$ region of the Virgo Cluster due to dark matter annihilation; however further analysis is needed to confirm or deny this claim. Additionally, the BCG in A2029 may also host an AGN, which would also make detection of gamma-ray emission from the cluster as a whole potentially difficult. Coma specifically is a good candidate for detection. Recent radio observations of Coma show diffuse radio emission in the form of a halo and a relic (\citealt{Brown2011}), indicating the existence of non-thermal processes throughout the cluster, which imply the existence of gamma rays, and as shown here star formation in the galaxies within Coma can also give significant gamma-ray emission.

\subsection{Prospects for Detection with Ground-based Gamma-ray Telescopes}

Galaxy clusters have not yet been detected by ground based gamma-ray telescopes, such as H.E.S.S., MAGIC or VERITAS. Recent studies of the Perseus cluster by MAGIC did not yield a detection apart from the central AGN, NGC 1275 and a radio galaxy, IC 310 \citep{Aleksic2012}, However, the gamma-ray spectrum of NGC 1275 as measured by MAGIC drops off above 630 GeV, which means that Perseus is a good potential candidate for detection in this energy range. The lower limit gamma-ray flux of Perseus above a TeV, about $1\times10^{-19}$ cm$^{-2}$ s$^{-1}$, is approximately 6 orders of magnitude smaller than the upper limits reported in \citet{Aleksic2012}, and well below the sensitivity limits of current atmospheric cherenkov telescopes. The gamma-ray emission associated with star formation in Perseus is therefore unlikely to be detected by ground-based instruments.

We also investigated whether the next-generation ground-based telescopes, such as the Cherenkov Telescope Array (CTA), would be able to produce a detection of galaxy clusters. As described in \citet{Actis2011}, the effective area of CTA is typically the limiting factor for a given observing time, usually about 25-50 hours. If we assume a constant power law spectrum for the gamma-ray emission from star-forming galaxies in clusters with a photon index $\alpha_{\gamma}=2.2$, we can estimate the lower limits on the flux from the clusters in our sample from 100 GeV to 10 TeV, which is the primary target energy range for CTA. The cluster with the largest flux in this range is A2029 with $2.6\times 10^{-15}$ cm$^{-2}$ s$^{-1}$; Virgo then Coma have the next highest fluxes at $6.7\times 10^{-16}$ cm$^{-2}$ s$^{-1}$ and $1.5\times 10^{-16}$ cm$^{-2}$ s$^{-1}$, respectively. A1367 and Hydra also have predicted fluxes that are a factor of $\sim2$ lower than Coma's. Using the flux of Virgo, the required effective area of CTA for 50-hour observation would be about 8 km$^{2}$; CTA will ideally cover at tens of km$^2$, at least. These estimates assume no spectral breaks at higher energies; as described in \citet{Bechtol2011}, starburst galaxies appear to have a spectrum described by a single power-law but the spectra of local galaxies such as the Milky Way are better described with an exponential cutoff or broken power-law. We therefore take these very-high-energy fluxes to be optimistic predictions. We conclude that the performance of CTA may allow for the detection of gamma rays from star formation in galaxy clusters for potentially several clusters.

\section{CONCLUSIONS}

We calculated lower limits on the gamma-ray emission from galaxy clusters considering only cluster member galaxies with active star formation using observed IR and radio luminosities for selected, nearby massive clusters. Employing the relationships derived in \citet{Bechtol2011} for \textit{Fermi}-LAT detected galaxies, we converted IR and radio cluster luminosities into gamma-ray luminosities. Several clusters have lower limits on their gamma-ray emission that are within about an order of magnitude of the upper limits based on the \textit{Fermi}-LAT non-detections from \citet{Ackermann2010c}, implying that star formation could contribute at the level of 10\% or more to cluster gamma-ray emission. Several clusters also have lower limits that are within a factor of a few of the 10-year sensitivity limits of the \textit{Fermi}-LAT; the best candidate clusters for detection by \textit{Fermi}-LAT based on these lower limits are A2029, Virgo, Coma, A1367, and Hydra. CTA will also likely be able to detect clusters such as A2029, Virgo and Coma for anticipated instrumental performance and design. Star formation may thus be a significant source of gamma-ray emission for some galaxy clusters.   
 
\acknowledgments
This work is partly supported by NASA grants NNX09AT96G and NNX09AT83G. SP acknowledges support from an Outstanding Junior Investigator Award from the Department of Energy, DE-FG02-04ER41286.

\bibliography{grs_from_sf}
\clearpage

\begin{table}

\begin{center}
  Table 1\\
  IR and Radio Luminosities
\\
\begin{tabular}{  c c c c c c  } \hline
  Name & log(L$_{8-1000 \mu m})$ & & log(L$_{1.4 \textrm{GHz}}$) & & D$_{\textrm{L}}$ \\ 
  & (L$_{\sun}$) & Ref. & (W Hz$^{-1}$) & Ref. & (Mpc) \\ \hline 
  Coma (LF)$^{a}$ & 11.75 & 2 & ... & ... & 100 \\
  Coma & 11.32 & 1 & 23.16 & 1 & 100 \\ 
  AWM7 & 10.80 & 1 & 22.28 & 1 & 69.2 \\
  Perseus & 10.53 & 1 & 22.05 & 1 & 72.3 \\
  Hydra & 11.10  & 1 & 22.68 & 1 &  57.1 \\ 
  A1367 & 11.48 & 1 & 23.25 & 1 & 96.6 \\
  Virgo & 10.86 & 1 & 22.42 & 1 & 19.4 \\
  Bullet & 12.63 & 3 & ... & ... & 1479 \\
  Ophiuchus$^{b}$ & ... & ... & 22.68 & 4 & 118\\
  A2029$^{b}$ & ... & ... & 24.86 & 4 & 339 \\
  A2142$^{b,c}$ & ... & ... & 22.68 & 4 & 401 \\ \hline
\end{tabular}
\end{center}

\textbf{Notes.} D$_{\textrm{L}}$ is luminosity distance, retrieved from the NASA/IPAC Extragalactic Database (NED). The NASA/IPAC Extragalactic Database (NED) is operated by the Jet Propulsion Laboratory, California Institute of Technology, under contract with the National Aeronautics and Space Administration.

($^{a}$) LF is luminosity function. See text for details.

($^{b}$) The data for these clusters is for BCGs only, not the full clusters.

($^{c}$) The radio luminosity for the BCG of A2142 is an upper limit.

\textbf{References.} (1) \citet{Reddy2004}; (2) \citet{Bai2006}; (3) \citet{Chung2010}; (4) \citet{McDonald2011}.
\end{table}

\begin{table}
\begin{center}
  Table 2 \\ 
  Upper and Lower Limits on Gamma-ray Luminosities
  \\
\begin{tabular}{  c c c c  } \hline
  Cluster & log(L$_{0.1-100 \textrm{GeV}}$) & log(L$_{0.1-100 \textrm{GeV}}$) & log(L$_{0.1-100 \textrm{GeV}}$) \\ 
 Name & (from L$_{8-1000 \mu\textrm{m}}$; $\pm$0.25 dex) & (from L$_{1.4 \textrm{GHz}}$; $\pm$0.19 dex) & \\
  & lower limit & lower limit & upper limit \\ \hline 
  Coma (LF)$^{a}$ & 41.07 & ... & 42.55 \\
  Coma & 40.50 & 41.07 & 42.55 \\ 
  AWM7 & 40.01 & 40.17 & 42.15 \\
  Perseus & 39.73 & 39.93  & 43.55 \\
  Hydra & 40.28 & 40.55 & 41.74 \\ 
  A1367 & 40.70 & 41.19  & 42.07 \\
  Virgo & 40.04 & 40.29 & 41.61 \\
  Bullet & 42.06 & ... & 44.66 \\
  Ophiuchus$^{b}$ & ... & 40.66 & 43.45\\
  A2029$^{b}$ & ... & 43.06 & 43.46 \\
  A2142$^{b}$ & ... & 40.66 & 43.54 \\ \hline
\end{tabular}
\end{center}

\textbf{Notes.} Luminosities are reported in erg s$^{-1}$. The uncertainties in the lower limit luminosities are from \citet{Bechtol2011} and are a measure of the dispersion in the relationships used to calculate L$_{0.1-100 \textrm{GeV}}$. Upper limits on gamma-ray luminosity (column 4) assume a power law spectrum with $\alpha_{\gamma}=2$ in converting from the $0.2-100$ GeV energy band, as in \citealt{Ackermann2010c}, to $0.1-100$ GeV for direct comparision with lower limits estimated from IR and radio luminosities.

($^{a}$) LF is luminosity function. See text for details.

($^{b}$) The data for these clusters is for BCGs only, not the full clusters.  

\end{table}

\clearpage
\normalsize
\begin{figure}
\includegraphics{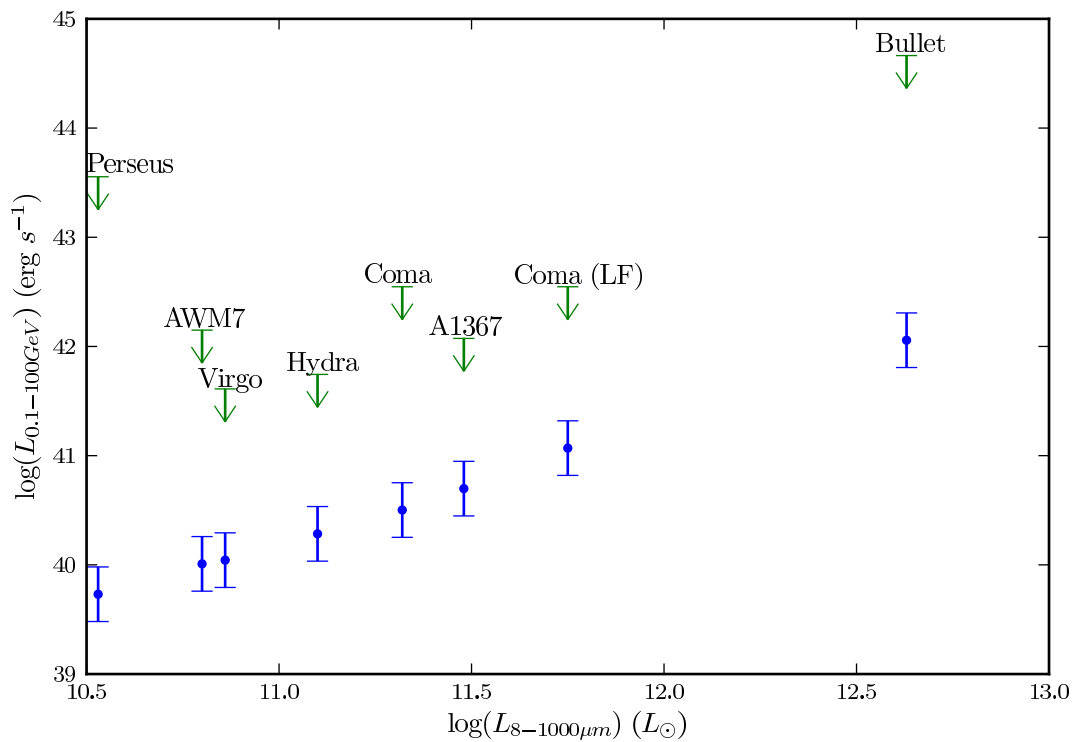}
\caption{Gamma-ray luminosity versus total infrared luminosity. Blue points correspond to lower limits on gamma-ray luminosity from star formation, calculated using relations in \citet{Bechtol2011}. The uncertainties are reported in \citet{Bechtol2011} and are a measure of the dispersion in the L$_{0.1-100\textrm{GeV}}$--L$_{8-1000 \mu\textrm{m}}$ relationship. Green arrows correspond to upper limits from \citet{Ackermann2010c}.}
\end{figure}

\clearpage
\normalsize
\begin{figure}
\includegraphics{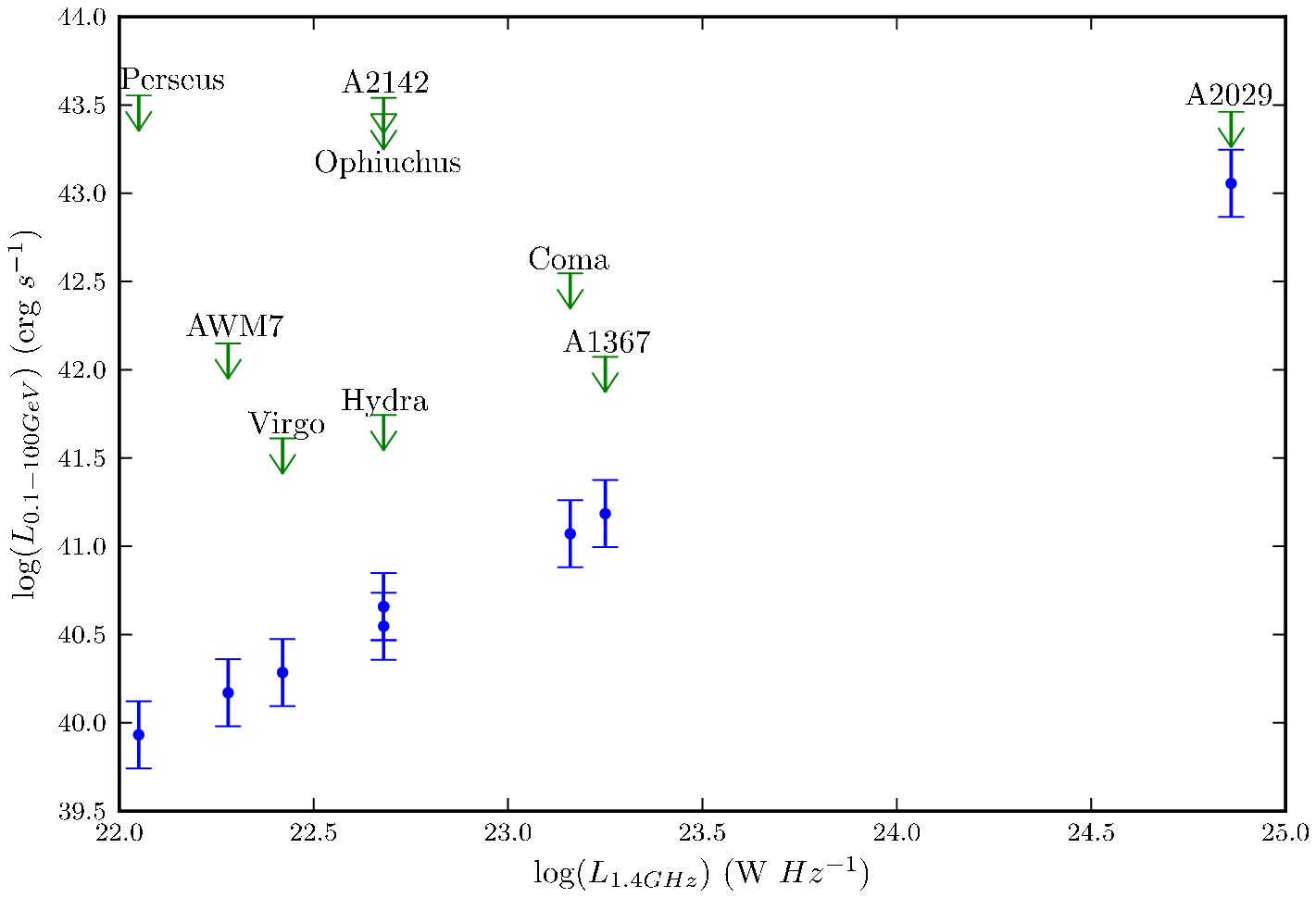}
\caption{Gamma-ray luminosity versus radio luminosity. Blue points correspond to lower limits on gamma-ray luminosity from star formation, calculated using relations in \citet{Bechtol2011}. The uncertainties are reported in \citet{Bechtol2011} and are a measure of the dispersion in the L$_{0.1-100\textrm{GeV}}$--L$_{1.4 \textrm{GHz}}$ relationship. Green arrows correspond to upper limits from \citet{Ackermann2010c}.}
\end{figure}

\end{document}